\documentclass[sigconf,natbib=true,anonymous=false]{acmart}

\usepackage{amssymb}
\usepackage{array}
\usepackage{url}
\usepackage{dblfloatfix}
\usepackage{booktabs}
\usepackage{amsmath}
\usepackage{tabularx}
\usepackage{colortbl}

\usepackage{graphics}
\usepackage{epsfig}
\usepackage{multirow}
\usepackage{flushend}

\usepackage{url}
\usepackage{tikz}
\usepackage{enumitem}
\usepackage{multicol}

\usepackage{verbatim}
\usepackage{float}
\usepackage{subcaption}

\usepackage{algorithm}
\usepackage{algpseudocode}

\usepackage{marginnote}
\usepackage{xspace}
\usepackage{balance}

\usepackage{todonotes}
\usepackage{xcolor}

\newcommand{\squishlist}{
 \begin{list}{$\bullet$}
  { \setlength{\itemsep}{0pt}
     \setlength{\parsep}{1pt}
     \setlength{\topsep}{1pt}
     \setlength{\partopsep}{0pt}
     \setlength{\leftmargin}{1.5em}
     \setlength{\labelwidth}{1em}
     \setlength{\labelsep}{0.5em} } }

\newcommand{\squishend}{
  \end{list}  }


\author{Iain Mackie}
\affiliation{
  \institution{University of Glasgow}
}
\email{i.mackie.1@research.gla.ac.uk}

\author{Shubham Chatterjee}
\affiliation{
  \institution{University of Glasgow}
}
\email{shubham.chatterjee@glasgow.ac.uk}

\author{Jeffrey Dalton}
\affiliation{
  \institution{University of Glasgow}
}
\email{jeff.dalton@glasgow.ac.uk}


\copyrightyear{2023} 
\acmYear{2023} 
\setcopyright{acmlicensed}\acmConference[SIGIR '23]{Proceedings of the 46th International ACM SIGIR Conference on Research and Development in Information Retrieval}{July 23--27, 2023}{Taipei, Taiwan}
\acmBooktitle{Proceedings of the 46th International ACM SIGIR Conference on Research and Development in Information Retrieval (SIGIR '23), July 23--27, 2023, Taipei, Taiwan}
\acmPrice{15.00}
\acmDOI{10.1145/3539618.3591992}
\acmISBN{978-1-4503-9408-6/23/07}

\renewcommand\footnotetextcopyrightpermission[1]{} 

\begin{document}
\fancyhead{}

\title{Generative Relevance Feedback with Large Language Models}

\begin{abstract}

Current query expansion models use pseudo-relevance feedback to improve first-pass retrieval effectiveness; however, this fails when the initial results are not relevant. 
Instead of building a language model from retrieved results, we propose Generative Relevance Feedback (GRF) that builds probabilistic feedback models from long-form text generated from Large Language Models. 
We study the effective methods for generating text by varying the zero-shot generation subtasks: queries, entities, facts, news articles, documents, and essays. 
We evaluate GRF on document retrieval benchmarks covering a diverse set of queries and document collections, and the results show that GRF methods significantly outperform previous PRF methods. 
Specifically, we improve MAP between 5-19\% and NDCG@10 17-24\% compared to RM3 expansion, and achieve the best R@1k effectiveness on all datasets compared to state-of-the-art sparse, dense, and expansion models.

\end{abstract}

\maketitle

%%%%%%%%%%%%%%%%%%%%%%%%%%%%%%%%%%%%%%%%%%%%%%%%%%%%%%%%%%%%%%%%%%%%%%%
%%%%%%%%%%%%%%%%%%%%%%%%%%%% NEW SECTION %%%%%%%%%%%%%%%%%%%%%%%%%%%%%%
%%%%%%%%%%%%%%%%%%%%%%%%%%%%%%%%%%%%%%%%%%%%%%%%%%%%%%%%%%%%%%%%%%%%%%%
\section{Introduction}
\label{sec:intro}

% -- lexical mismatch -- 
Recent advances in Large Language Models (LLMs) such as GPT-3 \cite{brown2020gpt3}, PaLM~\cite{chowdhery2022palm}, and ChatGPT demonstrate new capabilities to generate long-form fluent text. In addition, LLMs are being combined with search engines, including BingGPT or Bard, to create summaries of search results in interactive forms. In this work, we use these models not to generate end-user responses but as input to the core retrieval algorithm. 

The classical approach to address the vocabulary mismatch problem~\cite{belkin1982ask} is query expansion using Pseudo-Relevance Feedback (PRF)~\cite{abdul2004umass, zhai2001model, metzler2007latent, metzler2005markov}, where the  query is expanded using terms from the top-$k$ documents in a feedback set. This feedback set is obtained using a first-pass retrieval, and the expanded query is then used for a second-pass retrieval. While query expansion with PRF often improves recall, its effectiveness hinges on the quality of the first-pass retrieval. Non-relevant results in the feedback set introduce noise and may pull the query off-topic.

\begin{figure}[h!]
    \centering
    \includegraphics[scale=0.16]{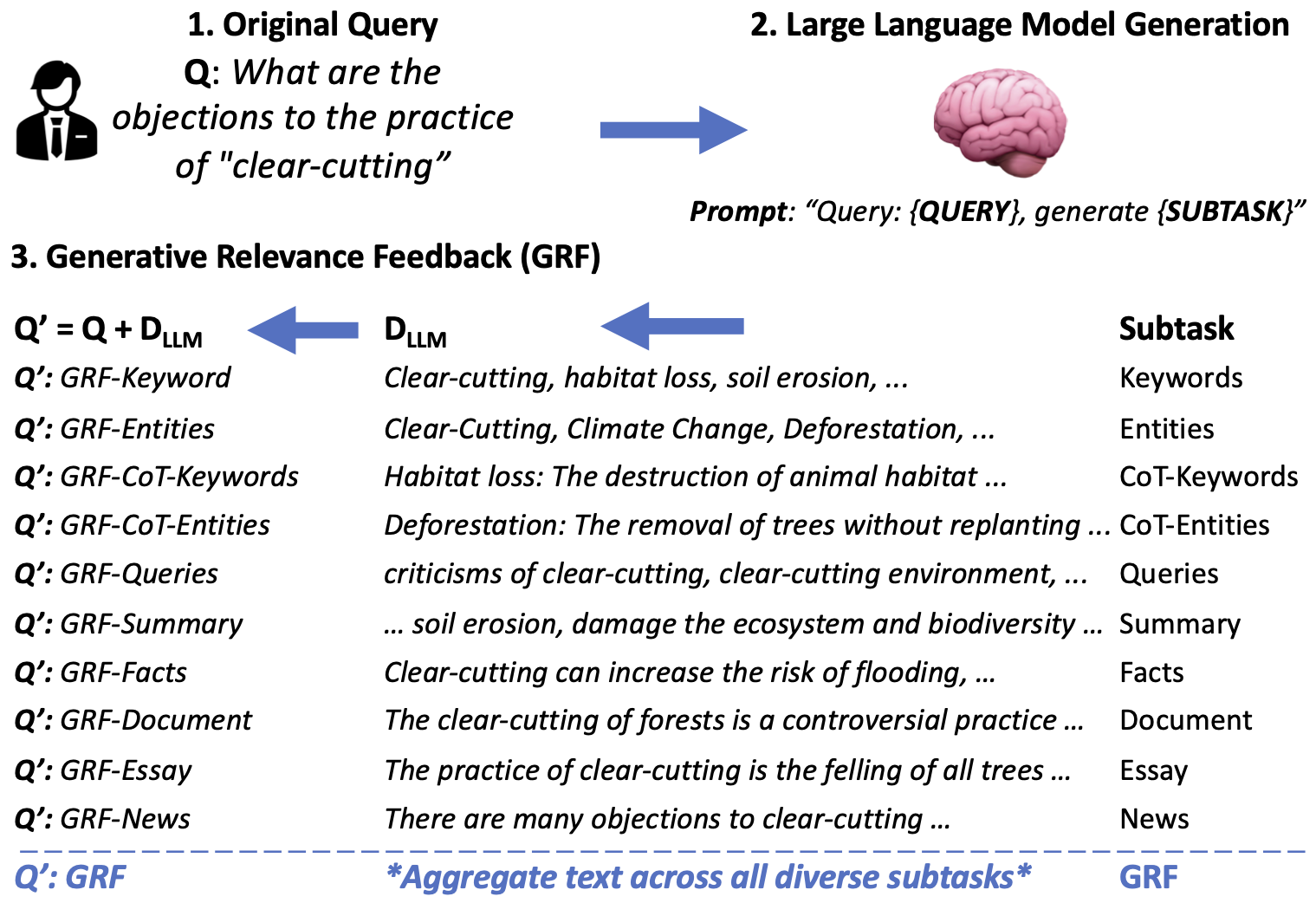}
    \caption{GRF uses diverse LLM-generate text content for relevance feedback to contextualise the query.
    }
    \label{img:grf-overview}
\end{figure}

% Solution / Setup - GRF
To address this problem, we propose Generative Relevance Feedback (GRF) that uses LLMs to generate text independent of first-pass retrieval. 
Figure~\ref{img:grf-overview} shows how we use an LLM to generate diverse types of query-specific text, before using these ``generated documents'' as input for proven query expansion models~\cite{abdul2004umass}.
We experiment using the following types of generated text: keywords, entities, chain-of-thought reasoning, facts, news articles, documents, and essays. Furthermore, we find that combining text across all generation subtasks results in 2-7\% higher MAP versus the best standalone generation. 

We evaluate GRF\footnote{ \href{https://drive.google.com/drive/folders/1LWGTvXGatrAbwbDahYkraK-nim2O9eyN?usp=sharing}{\textbf{Prompts and generated data for reproducibility: link}}} on four established document ranking benchmarks (see Section \ref{subsec:Datasets}) and outperform several state-of-the-art sparse~\cite{robertson1994some, abdul2004umass}, dense~\cite{naseri2021ceqe, wang2022colbert, li2022improving}, and learned sparse~\cite{lassance2023naver} PRF models. We find that long-form text generation (i.e. news articles, documents, and essays) is 7-14\% more effective as a feedback set compared to shorter texts (i.e. entities and keywords). Furthermore, the closer the generation subtask is to the style of the target dataset (i.e. news generation for newswire corpus or document generation for web document corpus), the more effective GRF is. Lastly, combining text across all generation subtasks results in 2-7\% improvement in MAP over the best standalone generation subtask.

The contributions of this work are:

\begin{itemize}[leftmargin=*]

\item We propose GRF, a generative relevance feedback approach which builds a relevance model using text generated from an LLM. 

\item 
We show LLM generated long-form text in the style of the target dataset is the most effective. Furthermore, combing text across multiple generation subtasks can further improve effectiveness. 

\item We demonstrate that GRF improves MAP between 5-19\% and NDCG@10 between 17-24\% compared to RM3 expansion, and achieves the best Recall@1000 compared to state-of-the-art sparse, dense and learned sparse PRF retrieval models.

\end{itemize}

\section{Related Work}
\label{sec:related-word}

\textbf{Query Expansion}:
% \label{sec:rw-qe}
Lexical mismatch is a crucial issue in information retrieval, whereby a user query fails to capture their complete information need~\cite{belkin1982ask}. Query expansion methods~\cite{rocchio1971relevance} tackle this problem by incorporating terms closer to the user's intended meaning. One popular technique for automatic query expansion is pseudo-relevance feedback (PRF), where the top $k$ documents from the initial retrieval are assumed to be relevant.
For example, Rocchio~\cite{rocchio1971relevance}, KL expansion~\cite{zhai2001model}, relevance modelling~\cite{metzler2005markov}, LCE~\cite{metzler2007latent}, and RM3 expansion~\cite{abdul2004umass}.
Additionally, we have seen approaches that expand queries with KG-based information~\cite{meij2010conceptual, xiong2015query, Xu2009query, dalton2014entity} or utilize query-focused LLM vectors for query expansion~\cite{naseri2021ceqe}.

Recent advancements in dense retrieval~\cite{khattab2020colbert, lin2020distilling, xiongapproximate} have led to the development of vector-based PRF models~\cite{li2022improving}, such as ColBERT PRF~\cite{wang2022colbert}, ColBERT-TCT PRF~\cite{yu2021improving}, and ANCE-PRF~\cite{yu2021improving}. Furthermore, SPLADE~\cite{formal2021splade} is a neural retrieval model that uses BERT and sparse regularization to learn query and document sparse expansions. Recent work has leveraged query expansion with PRF of learned sparse representations~\cite{lassance2023naver}. 
%By contrast, our approach builds a term-based relevance model using generated text content from LLMs, which is exposed through our generation subtasks. 
Unlike prior work, GRF does not rely on pseudo-relevance feedback, instead generating relevant text context for query expansion using LLMs.

\noindent \textbf{LLM Query Augmentation}
% \label{sec:rw-tg}
The emergence of LLMs has shown progress across many different aspects of information retrieval~\cite{yates2021pretrained}.
This includes using LLMs to change the query representation, such as query generation and rewriting~\cite{jeronymo2023inpars, nogueira2019doc2query, zamani2020generating, wu-etal-2022-conqrr, samarinas2022revisiting, macavaney2021intent5}, context generation~\cite{hyde, liu2022query}, and query-specific reasoning~\cite{pereira2023visconde, ferraretto2023exaranker}.
For example, \citet{nogueira2019doc2query} fine-tune a T5 model to generate queries for document expansion for passage retrieval.
More recent work by \citet{bonifacio2022inpars} shows that GPT3 can be effectively leveraged for few-short query generation for dataset generation.
Furthermore, LLMs have been used for conversational query re-writing~\cite{wu-etal-2022-conqrr} and generating clarifying questions~\cite{zamani2020generating}.  

We have also seen facet generation using T5~\cite{macavaney2021intent5} and GPT3~\cite{samarinas2022revisiting} to improve the relevance or diversity of search results.
While in QA, \citet{liu2022query} sample various contextual clues from LLMs and augment and fuse multiple queries.
For passage ranking, HyDe~\cite{hyde} uses InstructGPT~\cite{ouyang2022training} to generate hypothetical document embeddings and use Contriever~\cite{izacard2021contriever} for dense retrieval.
Lastly, works have shown LLM generation used for query-specific reasoning~\cite{pereira2023visconde, ferraretto2023exaranker} to improve ranking effectiveness.
Our approach differs from prior LLM augmentation approaches as we use LLMs to generate long-form text to produce a probabilistic expansion model to tackle query-document lexical mismatch.

\section{Generative Relevance Feedback}
\label{sec:method}

% -- Overview -- 
Generative Relevance Feedback (GRF) tackles query-document lexical mismatch using text generation for zero-shot query expansion.
Unlike traditional PRF approaches for query expansion~\cite{abdul2004umass, metzler2005markov, metzler2007latent}, GRF is not reliant on first-pass retrieval effectiveness to find useful terms for expansion.
Instead, we leverage LLMs~\cite{brown2020language} to generate zero-shot relevant text content.

% -- Relevance model and Generation subtasks -- 
We build upon prior work on Relevance Models~\cite{abdul2004umass} to incorporate the probability distribution of the terms generated by our LLM.
This approach enriches the original query with useful terms from diverse generation subtasks, including keywords, entities, chain-of-thought reasoning, facts, news articles, documents, and essays.
We find that the most effective query expansions are: (1) long-form text generations and (2) text content closer in style to the target dataset. 
In essence, we show that LLMs can effectively generate zero-shot text context close to the target relevant documents.

% -- Full GRF -- 
Lastly, we propose our full GRF method that combines text content across all generation subtasks.
The intuition behind this approach is that if terms are used consistently generated across subtasks (i.e. within the entity, fact, and news generations), then these terms are likely useful for expansion.
Additionally, multiple diverse subtasks also help expose tail knowledge or uncommon synonyms helpful for retrieval.   
We find this approach is more effective than any standalone generation subtasks.

\subsection{GRF Query Expansion}
\label{sec:text-gen}

For a given query $Q$, Equation \ref{eq:prob-grf} shows how $P_{GRF}(w|R)$ is the probability of a term, $w$, being in a relevant document, $R$.  
Similar to RM3~\cite{abdul2004umass}, GRF expansion combines the probability of a term given the original query $P(w|Q)$ with the probability of a term within our LLM-generated document, $P(w|D_{LLM})$, which we assume is relevant.
$\beta$ (original query weight) is a hyperparameter to weigh the relative importance of our generative expansion terms.
Additionally, $\theta$ (number of expansion terms) is a hyperparameter with $W_{\theta}$ being the set of most probable LLM-generated terms.

\begin{equation}
  \centering
   \label{eq:prob-grf}
  P_{GRF}(w|R) = \beta P(w|Q) + \begin{cases}
    (1 - \beta) P(w|D_{LLM}), & \text{if $w \in W_{\theta}$}.\\
    0, & \text{otherwise}.
  \end{cases}
\end{equation}

\subsection{Generation Subtasks}
\label{sec:query-exp}

We study how LLMs can generate relevant text, $D_{LLM}$, across diverse generation subtasks for GRF expansion.
% This allows us to analyse how these different subtasks perform on different target datasets in Section \ref{sec:results}.
The 10 query-specific generation subtasks are:

\begin{itemize}[leftmargin=*]

\item \textbf{Keywords (64 tokens)}: Generates a list of the important words or phrases for the topic, similar to facet generation~\cite{macavaney2021intent5, samarinas2022revisiting}. 
 
\item \textbf{Entities (64 tokens)}: Generates a list of important concepts or named entities, similar to KG-based expansion approaches~\cite{dalton2014entity}. 

\item \textbf{CoT-Keywords (256 tokens)}: Generate chain-of-thought (CoT)~\cite{weichain} reasoning to explain ``why'' a list of keywords are relevant.

\item \textbf{CoT-Entities (256 tokens)}: Generate CoT reasoning to explain ``why'' a list of entities are relevant.

\item \textbf{Queries (256 tokens)}: Generate a list of queries based on the original query, similar to~\cite{bonifacio2022inpars}.

\item \textbf{Summary (256 tokens)}: Generate a concise summary (or answer) to satisfy the query. 

\item \textbf{Facts}: Generate a knowledge-intensive list of text-based facts on the topic, which is close to~\cite{liu2022query}.

\item \textbf{Document (512 tokens)}: Generate a relevant document based on the query closest to a long-form web document.

\item \textbf{Essay (512 tokens)}: Generate a long-form essay-style response.

\item \textbf{News (512 tokens)}: Generate text in the style of a news article.

\end{itemize}

The full \textbf{GRF} expansion model concatenates text generated across all subtasks to produce $D_{LLM}$. We then calculate $P(w|D_{LLM})$ using this aggregated text, as outlined above.
Section \ref{sec:results} shows that the combination using the text across all types is most effective.

\begin{table*}[b!] 
% \small
\caption{GRF with different generation subtasks. Significant improvements against BM25+RM3 (``+'') and best system (\textit{bold}).}
\tabcolsep=0.04cm
\label{tab:content}
\begin{tabular}{l|ccc|ccc|ccc|ccc|}
\cline{2-13}
                                   & \multicolumn{3}{c|}{Robust04 -Title}                 & \multicolumn{3}{c|}{CODEC}                               & \multicolumn{3}{c|}{DL-19}                                   & \multicolumn{3}{c|}{DL-20}                                   \\ \cline{2-13} 
                                   & NDCG@10            & MAP            & R@1k           & NDCG@10            & MAP                & R@1k           & NDCG@10            & MAP                & R@1k               & NDCG@10            & MAP                & R@1k               \\ \hline
\multicolumn{1}{|l|}{BM25}         & 0.445              & 0.252          & 0.705          & 0.316              & 0.214              & 0.783          & 0.531              & 0.335              & 0.703              & 0.546              & 0.413              & 0.811              \\
\multicolumn{1}{|l|}{BM25+RM3}     & 0.451              & 0.292          & 0.777          & 0.326              & 0.239              & 0.816          & 0.541              & 0.383              & 0.745              & 0.513              & 0.418              & 0.825              \\ \hline
\multicolumn{1}{|l|}{GRF-Keywords}     & 0.435              & 0.252          & 0.717          & 0.327              & 0.218              & 0.748          & 0.565              & 0.377              & 0.749              & 0.554              & 0.435              & 0.822              \\
\multicolumn{1}{|l|}{GRF-Entities}     & 0.452              & 0.252          & 0.698          & 0.341              & 0.216              & 0.750          & 0.531              & 0.363              & 0.741              & 0.544              & 0.414              & 0.824              \\
\multicolumn{1}{|l|}{GRF-CoT-Keywords} & 0.436              & 0.248          & 0.704          & 0.327              & 0.239              & 0.774          & 0.550              & 0.382              & 0.748              & 0.542              & 0.423              & 0.817              \\
\multicolumn{1}{|l|}{GRF-CoT-Entities} & 0.450              & 0.252          & 0.714          & 0.355              & 0.243              & 0.789          & 0.563              & 0.389              & 0.757              & 0.552              & 0.430              & 0.832              \\
\multicolumn{1}{|l|}{GRF-Queries}      & 0.450              & 0.257          & 0.710          & 0.347              & 0.233              & 0.773          & 0.551              & 0.367              & 0.760              & 0.568              & 0.439              & 0.851              \\
\multicolumn{1}{|l|}{GRF-Summary}      & 0.491$^+$          & 0.277          & 0.730          & 0.398$^+$          & 0.260              & 0.796          & 0.577              & 0.414              & 0.761              & 0.585$^+$          & 0.472$^+$          & 0.865              \\
\multicolumn{1}{|l|}{GRF-Facts}        & 0.501$^+$          & 0.284          & 0.744          & 0.353              & 0.255              & 0.795          & 0.569              & 0.401              & 0.769              & 0.583$^+$          & 0.459$^+$          & 0.871              \\
\multicolumn{1}{|l|}{GRF-Document}     & 0.480$^+$          & 0.276          & 0.728          & 0.376$^+$          & 0.265              & 0.795          & 0.618$^+$          & 0.428$^+$          & 0.787$^+$          & 0.589$^+$          & 0.476$^+$          & 0.872              \\
\multicolumn{1}{|l|}{GRF-Essay}        & 0.494$^+$          & 0.284          & 0.736          & \textbf{0.405$^+$}          & 0.270$^+$          & 0.803          & 0.609$^+$          & 0.421$^+$          & 0.779$^+$          & 0.551              & 0.440              & 0.859              \\
\multicolumn{1}{|l|}{GRF-News}         & 0.501$^+$          & 0.287          & 0.745          & 0.398$^+$          & 0.270$^+$          & 0.828          & 0.609              & 0.409              & 0.777              & 0.578$^+$          & 0.457              & 0.853              \\ \hline
\multicolumn{1}{|l|}{GRF}          & \textbf{0.528$^+$} & \textbf{0.307} & \textbf{0.788} & \textbf{0.405$^+$} & \textbf{0.285$^+$} & \textbf{0.830} & \textbf{0.620$^+$} & \textbf{0.441$^+$} & \textbf{0.797$^+$} & \textbf{0.607$^+$} & \textbf{0.486$^+$} & \textbf{0.879$^+$} \\ \hline
\end{tabular}
\end{table*}

\section{Experimental Setup}
\label{sec:exp-setup}

\subsection{Datasets}
\label{subsec:Datasets}

\subsubsection{Retrieval Corpora}
  
\textbf{TREC Robust04}~\cite{Voorhees_TREC2004_robust} was created to investigate methods targeting poorly performing topics. 
This dataset comprises 249 topics, containing short keyword ``titles'' and longer natural-language "descriptions" queries. 
Relevance judgments are over a newswire collection of 528k long documents (TREC Disks 4 and 5), i.e. FT, Congressional Record, LA Times, etc.

%\noindent
\textbf{CODEC}~\cite{mackie2022codec} is a dataset that focuses on the complex information needs of social science researchers.
Domain experts (economists, historians, and politicians) generate 42 challenging essay-style topics.
CODEC has a focused web corpus of 750k long documents, which includes news (BBC, Reuters, CNBC etc.) and essay-based web content (Brookings, Forbes, eHistory, etc.).

%\noindent 
\textbf{TREC Deep Learning (DL) 19/20}~\cite{craswell2020overview, craswell2021overview} builds upon the MS MARCO web queries and documents~\cite{nguyen2016ms}. 
The TREC DL dataset uses NIST annotators to provide judgments pooled to a greater depth, containing 43 topics for DL-19 and 45 topics for DL-20. Both query sets are predominately factoid-based~\cite{mackie2021deep}.

\subsubsection{Indexing and Evaluation} For indexing we use Pyserini version 0.16.0~\cite{lin2021pyserini}, removing stopwords and using Porter stemming. 
We use cross-validation and optimise R@1k on standard folds for Robust04~\cite{huston2014parameters} and CODEC~\cite{mackie2022codec}. On DL-19, we cross-validated on DL-20 and use the average parameters zero-shot on DL-19 (and vice versa for DL-20).
We assess the system runs to a run depth of 1,000. 
With GRF being an initial retrieval model, recall-oriented evaluation is important, such as Recall@1000 and MAP to identify relevant documents.
We also analyse NDCG@10 to show precision in the top ranks. 
We use ir-measures for all our evaluations~\cite{macavaney2022streamlining} and a 95\% confidence paired-t-test for significance.

\subsection{GRF Implementation}

\textbf{LLM Generation.} For our text generation we use the GPT3 API~\cite{brown2020language}.
Specifically, we use the \texttt{text-davinci-002} model with parameters: \textit{temperature} of 0.7, \textit{top\_p} of 1.0, \textit{frequency\_penalty} of 0.0, and \textit{presence\_penalty} of 0.0.
We release all code, generation subtask prompts, generated text content and runs for reproducibility.

\textbf{Retrieval and Expansion} To avoid query drift, all GRF runs in the paper use a tuned BM25 system for the input initial run~\cite{robertson1994some}. 
We tune GRF hyperparameters: the number of feedback terms ($\theta$) and the interpolation between the original terms and generative expansion terms ($\beta$). The tuning methodology is the same as BM25 and BM25 with RM3 expansion to make the GRF directly comparable; see below for details.

\begin{table*}[h!]

\caption{GRF against state-of-the-art PRF models. Significant improvements against BM25+RM3 (``+'') and best system (\textit{bold}).}
\centering
\label{tab:sota}
\tabcolsep=0.04cm
\begin{tabular}{l|lll|lll|lll|lll|}
\cline{2-13}
                                   & \multicolumn{3}{c|}{Robust04 -Title}                     & \multicolumn{3}{c|}{CODEC}                               & \multicolumn{3}{c|}{DL-19}                                   & \multicolumn{3}{c|}{DL-20}                                   \\ \cline{2-13} 
                                   & nDCG@10            & MAP                & R@1k           & nDCG@10            & MAP                & R@1k           & nDCG@10            & MAP                & R@1k               & nDCG@10            & MAP                & R@1k               \\ \hline
\multicolumn{1}{|l|}{BM25+RM3}     & 0.451              & 0.292              & 0.777          & 0.326              & 0.239              & 0.816          & 0.541              & 0.383              & 0.745              & 0.513              & 0.418              & 0.825              \\
\multicolumn{1}{|l|}{CEQE-MaxPool} & 0.474              & \textbf{0.310$^+$} & 0.764          & -                  & -                  & -              & 0.518              & 0.378              & 0.746              & 0.473              & 0.396              & 0.841              \\
\multicolumn{1}{|l|}{SPLADE+RM3}   & 0.418              & 0.248              & 0.703          & 0.311              & 0.216              & 0.770          & 0.566              & 0.328              & 0.651              & 0.533              & 0.379              & 0.784              \\
\multicolumn{1}{|l|}{TCT+PRF}      & 0.493              & 0.274              & 0.684          & 0.358              & 0.239              & 0.757          & \textbf{0.670$^+$} & 0.378              & 0.684              & \textbf{0.618$^+$} & 0.442              & 0.784              \\
\multicolumn{1}{|l|}{ColBERT-PRF}  & 0.467              & 0.272              & 0.648          & -                  & -                  & -              & 0.668$^+$          & 0.385              & 0.625              & 0.615$^+$          & \textbf{0.489$^+$} & 0.813              \\ \hline
\multicolumn{1}{|l|}{GRF (Ours)}   & \textbf{0.528$^+$} & 0.307              & \textbf{0.788} & \textbf{0.405$^+$} & \textbf{0.285$^+$} & \textbf{0.830} & 0.620$^+$          & \textbf{0.441$^+$} & \textbf{0.797$^+$} & 0.607$^+$          & 0.486$^+$          & \textbf{0.879$^+$} \\ \hline
\end{tabular}
\end{table*}

\subsection{Comparison Methods}

\noindent \textbf{BM25}~\cite{robertson1994some}: 
Sparse retrieval method, we tune $k1$ parameter (0.1 to 5.0 with a step size of 0.2) and $b$ (0.1 to 1.0 with a step size of 0.1).

\noindent \textbf{BM25+RM3}~\cite{abdul2004umass}: 
For BM25 with RM3 expansion, we tune $fb\_terms$ (5 to 95 with a step of 5), $fb\_docs$ (5 to 50 with a step of 5), and $original\_query\_weight$ (0.2 to 0.8 with a step of 0.1). 

\noindent  \textbf{CEQE}~\cite{naseri2021ceqe}: Utilizes query-focused vectors for query expansion. We use the CEQE-MaxPool runs provided by the author. %, and for DL-19 and Dl-20 we use the CEQE runs provided by \cite{wang2023colbert}. 
%See respective papers for more details.

\noindent  \textbf{SPLADE+RM3}: We use RM3~\cite{abdul2004umass} expansion with SPLADE~\cite{formal2021splade}. We use \texttt{naver/splade-cocondenser-ensembledistil} checkpoint and Pyserini's~\cite{lin2021pyserini} ``impact'' searcher for max-passage aggregation. We tune $fb\_docs$ (5,10,15,20,25,30), $fb\_terms$ (20,40,60,80,100), and $original\_query\_weight$ (0.1 to 0.9 with a step of 0.1).

\noindent \textbf{TCT+PRF}:~\cite{Li2021PseudoRF} is a Roccio PRF approach using ColBERT-TCT~\cite{lin2021batch}. We employ a max-passage approach with \texttt{TCT-ColBERT-v2-HNP} checkpoint. We tune Roccio PRF parameters: $depth$ (2,3,5,7,10,17), $\alpha$ (0.1 to 0.9 with a step of 0.1), and $\beta$ (0.1 to 0.9 with a step of 0.1).  

\noindent \textbf{ColBERT+PRF}~\cite{wang2022colbert}: We use the runs provided by~\citet{wang2023colbert}, which use pyterrier framework~\cite{macdonald2021pyterrier} for ColBERT-PRF retrieval.

\section{Results \& Analysis}
\label{sec:results}

\subsection{RQ1: What generative content is most effective for query expansion?}

% --- table overview ---
Table \ref{tab:content} shows the effectiveness of generative feedback with varying units of text (Keywords-News) and our full hybrid method that uses text from all subtasks.
We test for significant improvements against BM25 with RM3 expansion, to ascertain whether our zero-shot generative feedback methods improve over RM3 expansion.

% --- keywords and entities + COT ---
Generation subtasks that target short text span or lists (Keywords, Entities, Keywords-COT, Entities-COT, and Queries) do not offer significantly improves over RM3 expansion.
Conversely, subtasks targeting long text generation (Summary, Facts, Document, Essay, News) significantly improve at least two datasets over RM3 expansions.
This indicates that more terms generated from the LLM provide a better relevance model, and increases MAP between 7-14\% when we compare these two categories.

% --- Target corpus ---
Furthermore, we find most effective generation subtasks are aligned with the style of the target dataset.
For example, Facts and News are the best standalone generation methods across all measures on Robust04, where the dataset contains fact-heavy topics and a newswire corpus.
Additionally, Essay and News are the best generation subtasks on CODEC across all measures, which aligns with its essay-style queries over a news (BBC, Reuters, CNBC, etc.) and essay-style (Brookings, Forbes, eHistory, etc.) corpus.
Lastly, Document is the best generation subtask across DL-19 and DL-20, aligning with MS Marcos web document collection.
Overall, this finding supports that LLM generative content in the styles of the target dataset is the most effective.

% --- Overall vs standalone---
Although we see significant improvements from some standalone generation subtasks, particularly NDCG@10 (15/40 subtasks across the datasets), the full GRF method is consistently as good if not better than any standalone subtask.
Specifically, GRF improves NDCG by 0.0-5.4\%, MAP by 2.1-7.0\% and R@1k by 0.2-5.8\% across the datasets.
This shows that combining LLM-generated text from various generation subtasks is a robust and effective method of relevance modelling.

% --- GRF vs RM3 ---
Lastly, these results show that GRF expansion from generated text is consistently better, often significantly, than RM3 expansion that uses documents from the target corpus.
Specifically, we find significant improvement on all measures across DL-19 and DL-20, NDCG@10 and MAP on CODEC, and NDCG@10 on Robust04 titles.
Although not included for space limitations, on Robust04 description queries, GRF shows significant improvements with an NDCG of 0.550, MAP of 0.318, and R@1k of 0.776.

% -- hardest queries ---
These results strongly support that LLM generation is an effective query expansion method without relying on first-pass retrieval effectiveness.
For example, we look at the hardest 20\% of Robust04 topics ordered by NDCG@10; we find that RM3 offers minimal uplift and only improves  NDCG@10 by +0.006, MAP by +0.008, and R@1k by +0.052. 
In contrast, GRF is not reliant on first-pass retrieval effectiveness, and GRF improves NDCG@10 by +0.145, MAP by +0.068, and R@1k by +0.165 (a relative improvement of +100-200\% on NDCG@10 and MAP).

\subsection{RQ2: How does GRF compare to state-of-the-art PRF models?}

Table \ref{tab:sota} shows GRF against state-of-the-art sparse, dense, and learned sparse PRF models across target datasets.
This allows us to directly compare GRF's unsupervised term-based queries against PRF methods that use more complex LLM-based embeddings.
We conduct significance testing against BM25 with RM3 expansion.

GRF has the best R@1k across all datasets and has comparable and often better effectiveness in the top ranks.
Specifically, on more challenging datasets, such as CODEC and Robust04 titles, GRF is the best system across all measures, except Robust04 titles MAP, which is 0.003 less than CEQE. 
Although not included for space limitations, GRF is also the most effective system on Robust04 descriptions across all measures.
GRF vastly outperforms dense retrieval and dense PRF on these challenging datasets, with a performance gap of 7-14\% on NDCG@20, 13-21\% MAP, and 10-22\% R@1k.

Dense retrieval has been shown to be highly effective on the more factoid-focused datasets, such as DL-19 and DL-20.
However, as well as the best R@1k, our unsupervised GRF queries have comparable NDCG@10 and MAP scores to dense PRF models.
This is juxtaposed to other sparse methods (BM25 and BM25 with RM3 expansions) or LLM expansion (CEQE), which have much poorer precision in the top ranks.
Overall, this supports that generative expansion is a highly effective initial retrieval method across various collections and query types.

\section{Conclusion}
\label{sec:conclusion}

To our knowledge, this is the first work to study the use of long-form text generated from large-language models for query expansion. We show that generating long-form text in news-like and essay-like formats is effective input for probabilistic query expansion approaches. The results on document retrieval on multiple corpora show that the proposed GRF approach outperforms models that use retrieved documents (PRF). The results show GRF improves MAP between 5-19\% and NDCG@10 between 17-24\% when compared to RM3 expansion, and achieves the best Recall@1000 compared to state-of-the-art PRF retrieval models. We envision GRF as one of the many new emerging methods that use LLM-generated content to improve the effectiveness of core retrieval tasks.

\section{Acknowledgements}
\label{sec:ack}

This work is supported by the 2019 Bloomberg Data Science Research Grant and the Engineering and Physical Sciences Research Council grant EP/V025708/1. 

\bibliographystyle{ACM-Reference-Format}
\balance
%\vfill\eject 
\bibliography{foo}

\end{document}